\documentclass[12pt,aps,prd,preprint,tightenlines,superscriptaddress,
   showpacs,nofootinbib]{revtex4}
\usepackage{hyperref}
\usepackage{graphicx}
\usepackage{epsfig}
\usepackage{amsmath, amsfonts,euscript,bbm}
\usepackage{ifthen}
\usepackage{graphicx}

\pdfoutput=1

\newcommand{\PRE}[1]{{#1}} 
\newcommand{\nn}{\nonumber}

\newcommand{\roughly}[1]{\mathrel{\raise.3ex\hbox{$#1$\kern-0.85em
\lower1ex\hbox{$\sim$}}}}

\def\be{\begin{equation}}
\def\beq\begin{equation}
\def\ee{\end{equation}}
\def\bea{\begin{eqnarray}}
\def\eea{\end{eqnarray}}
\def\nn{\nonumber}

\def\hf{\frac12}
\def\eqref#1{(\ref{#1})}

\def\UV{{\scriptscriptstyle U \kern-.1emV}}
\def\IR{{\scriptscriptstyle I\kern-.18em R}}

\newcommand{\Ket}[1]{\left|#1\right\rangle}

\newcommand{\Bra}[1]{\left\langle#1\right|}

\newcommand{\Expect}[1]{\left\langle #1 \right\rangle}

\def\gsim{\ \rlap{\raise 3pt \hbox{$>$}}{\lower 3pt \hbox{$\sim$}}\ }
\def\lsim{\ \rlap{\raise 3pt \hbox{$<$}}{\lower 3pt \hbox{$\sim$}}\ }

\def\T{{\rm T}}

\def\exd{{\rm d}}

\begin{document}

\preprint{UCI-TR-2011-14, Pi-cosmo-230}

\title{\PRE{\vspace*{1.3in}}
On the Decay of Massive Fields in de Sitter
\PRE{\vspace*{0.3in}}
}

\author{Dileep P.  Jatkar}
\affiliation{Harish-Chandra Research Institute, Chhatnag Road, Jhusi,
  Allahabad 211019 INDIA 
\PRE{\vspace*{.1in}}
}

\author{Louis Leblond}
\affiliation{Perimeter Institute, 31, Caroline Street, Waterloo,
  Ontario N2L 2Y5, Canada 
\PRE{\vspace*{.1in}}
}

\author{Arvind Rajaraman%
\PRE{\vspace*{.4in}}
}
\affiliation{Department of Physics and Astronomy, University of California, Irvine, CA  92697 USA
\PRE{\vspace*{.5in}}
}

\date{June 2011}

\begin{abstract}
  \PRE{\vspace*{.3in}} 
Interacting massive fields with $m> \frac{d H}{2}$ in $d+1$ dimensional de Sitter space are fundamentally unstable. Scalar fields in this mass range can decay to themselves. 
This process (which is kinematically forbidden in Minkowski space) can lead to an important change to the propagator and the physics of these fields. We compute this decay rate by doing a 1-loop computation for a massive scalar field with a cubic interaction.
We resum the 1-loop result by consistently solving the Schwinger-Dyson equations. We also perform an explicit resummation of all chain graphs in the case of the retarded propagator. The decay rate is exponentially suppressed for large $m/H$ and the flat space answer (vanishing decay rate) is reproduced in that limit. 
\end{abstract}

\maketitle

\section{Introduction}
Much work has been done in studying the quantum behavior of low mass
($m\ll H$) scalar fields in de Sitter space (dS).  In the inflationary paradigm,
quantum perturbations of these scalar fields give rise to
 the large scale structure that we see in the Universe today.
Comparatively 
less attention has been paid to massive fields with
$m\gg H$.  Clearly
in the limit of very large mass or vanishing Hubble rate, one should
recover the flat space answer.  
Recently, however, Krotov and Polyakov \cite{Krotov:2010ma} 
analyzed the correlation functions of these massive fields and found a surprising result. 
They found
that 
in the Poincar\'e patch, the  Keldysh propagator $F$ at 1-loop 
receives a 1-loop
logarithmic correction 
\be
F(k,\tau) \rightarrow F^0(k,\tau)\left( 1+ \sigma \ln k\tau\right) \; ,
\ee
where $F^0(k,\tau)$ is the tree level propagator, $k$ is the spatial momenta and 
$\tau$ is the conformal 
time. The loop correction naively diverges at late times $\tau \rightarrow 0^-$, 
which naively suggests a breakdown of perturbation theory. 

This 1-loop diagram had been analyzed before by Marolf and Morrison
\cite{Marolf:2010zp} working on the
sphere and analytically continuing to obtain the result in de
Sitter space.
By resuming all the 1PI diagrams they argued that this logarithmic correction
actually corresponds to an imaginary shift to the mass
\be
\mu \rightarrow \mu + i\sigma
\ee
where $\mu^2 = \frac{m^2}{H^2} - \frac{d^2}{4} > 0$ where $H$ is the Hubble constant and $d$ is the number of spatial dimensions. 
 In real space, the main effect of the resummation is to make the propagator decay
 faster at large distance
  \be
 G(Z) \rightarrow \frac{1}{Z^{d/2 -2\sigma}} ( A Z^{i\mu} + B Z^{-i\mu}) \; ,
\ee
where $Z$ is the geodesic distance, $A,B$ are specific coefficients and $\sigma$ will turn out to be real and negative.

In this paper we perform a resummation of the diagrams directly in Lorentzian signature.
We find that it is possible to resum all the bubble diagrams to show that
the 1-loop answer indeed leads to a (time independent) imaginary shift of the mass.  
The resummation is performed by consistently solving the Schwinger-Dyson equations for the system using
an ansatz for the full resummed propagators. 
In the case of the retarded propagator we directly carry out the resummation of all
bubble diagrams and show that the result is in agreement with the
solution of the Schwinger-Dyson equations. 
We also argue
that the same imaginary contribution can be found for quartic interactions. 
For low mass fields, very similar logs appear and these have already been
resummed using a variety of techniques (for recent work see \cite{Riotto:2008mv,Petri:2008ig,Rajaraman:2010zx,Burgess:2009bs, Giddings:2010nc, Rajaraman:2010xd,Garbrecht:2011gu}).
Our computations are all done in the Poincar\'e patch of de Sitter and
our resummation agrees with the 1PI resummation on the sphere
\cite{Marolf:2010zp} confirming that the two computations are
equivalent \cite{Higuchi:2010xt}. 

We note that this phenomenon may be related to another
fact: interacting massive fields in the
principal series ($m > \frac{d H}{2}$) appear to be fundamentally unstable.
Since conservation of energy does not hold in 
dS, it is possible for a particle to duplicate itself or even decay
into heavier ones \cite{Nacht}. Interestingly this is only true when $m> \frac{d H}{2}$.

To our knowledge the computation of this decay rate was first done by Bros, Esptein and Moschella in a series of papers \cite{Bros:2006gs, Bros:2008sq} (see also
\cite{Bros:2009bz, Bros:2010wa}).
In these papers, they computed the \emph{tree level} decay rate for a scalar field of mass $m$ with
cubic interaction $\lambda \phi^3$.
In flat space, the decay rate is
exactly zero simply because of kinematics. In de
Sitter space, however, they found that there is a non-zero but
exponentially small decay rate (for the principal series).  This is a reflection of the energy
non-conservation which opens up new decay channels.
In our case, we have found an imaginary shift in the mass, and it is plausible to relate
this to the decay rate found by \cite{Bros:2006gs, Bros:2008sq}.

We compute explicitly the decay rate (aka imaginary part to self-energy) and we show that it goes to zero roughly as $\lambda^2 \left(\frac{H}{m}\right)^n e^{-\pi m/H}$ 
in the large mass, low $H$ limit where the exponent $n$ depends on the number of dimensions. The decay rate goes to zero very
fast in the large $m/H$ limit and the flat space answer is recovered.
The decay of the Higgs (say with $m \sim 100 {\rm GeV}$) in our
Universe today (with $H_0 =  10^{-33} {\rm ev}$) is obviously very small. For $m \sim H$ the decay rate reaches a finite value of order $\Gamma \sim \lambda^2$.

This paper is
organised as follows. We set up the problem in section \S~\ref{setup} and compute
the 1-loop contribution in section \S~\ref{explicit} for all propagators. We then proceed to resum using
two different methods in section \S~\ref{resum}.  Finally, the decay rate is explicitly evaluated in section \S~\ref{decay}.
In an  appendix at the end, we show how a field with quartic interaction will have similar loop corrections.

\section{Interacting scalar field theory in de Sitter space}\label{setup}

\begin{figure} [ht]
\begin{center}
\includegraphics[width=0.7\textwidth,angle=0]{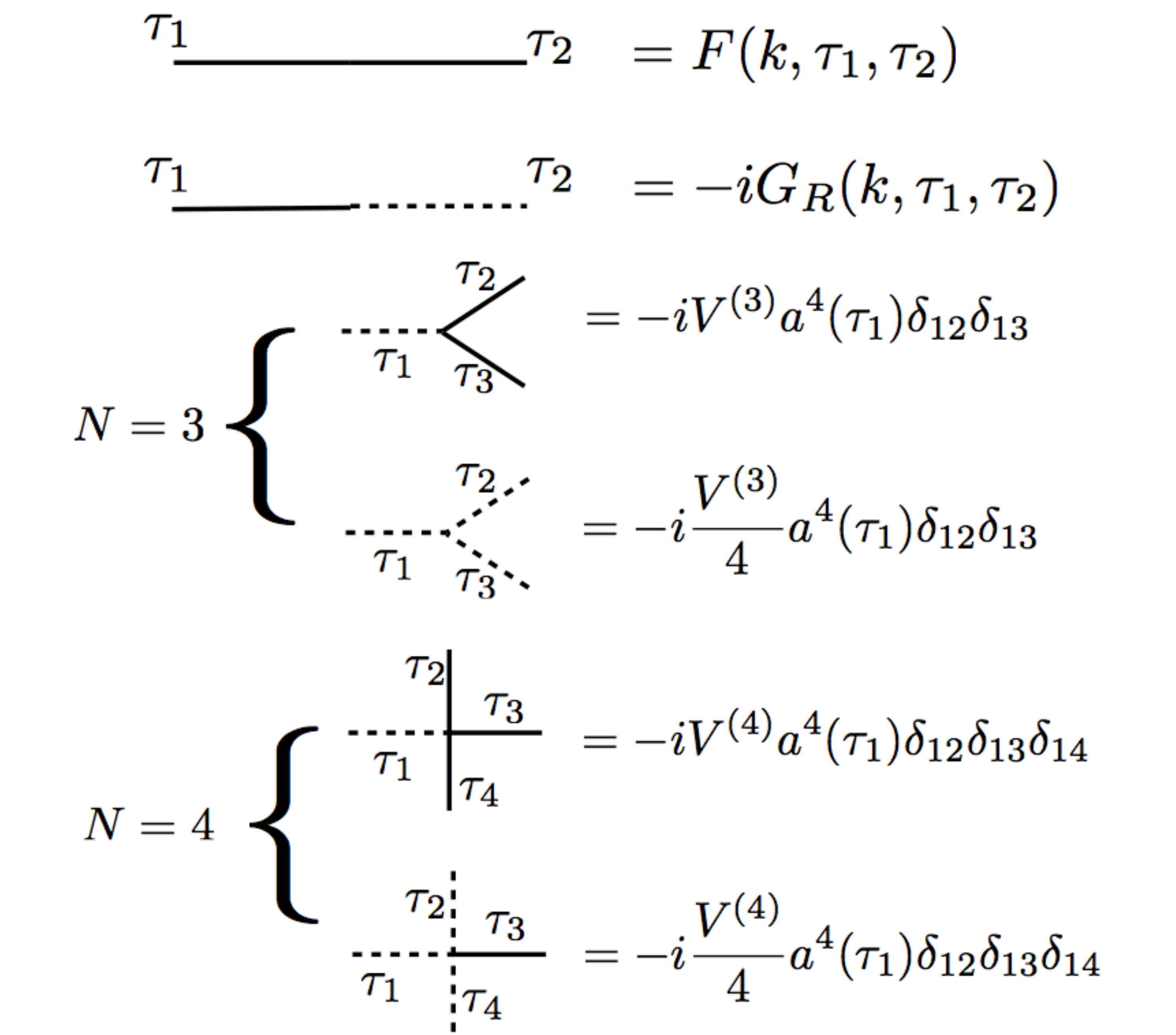}
\caption{Propagators and some vertices ($N=3$ and $N=4$) for a scalar with a potential of the form $\sum_N \frac{V^{(N)}}{N!} \phi^N$. Note that $G_A(k,\tau_1,\tau_2) = G_R(k,\tau_2,\tau_1)$. We use the shorthand notation $\delta_{12} = \delta(\tau_1-\tau_2)$. When a two-point function is attached to a vertex, the corresponding time must be integrated over. Internal spatial momenta must also be integrated over $\int \frac{d^dp}{(2\pi)^d}$. As usual, one must also divide by the symmetry factor for the diagram. At $N= 5$, there are 3 vertices, etc. The scale factor in dS with $H=1$ is just $a(\tau) = -1/\tau$.}.\label{prop}
\end{center}
\end{figure}

We are interested in calculating in-in correlation functions. 
\be
\Expect{Q(t)}  = \Bra{in} U^\dagger Q U \Ket{in}
\ee
where the evolution operators are
\begin{align}
U(t,t_0) &={\rm T\; exp}\left(-i\int_{t_0}^t dt' H_I(t')\right)\; , &  U^\dagger(t,t_0) =\overline{\T}\;{\rm exp}\left(i\int_{t_0}^t dt' H_I(t')\right) \; .
\end{align}
We take $t_0 =  t_{in}(1\mp i\epsilon) $ where the sign must be chosen accordingly to select free vacua in the far past and $H_I$ is the Hamiltonian in the interaction picture. One can compute the 1-loop contribution by directly expanding the evolution operator to the desired order. For the purpose of resummation, we find it useful to use Feynman rules that can be obtained by doubling the number of fields in the Keldysh-Schwinger formalism. For a scalar field with non-derivative interactions of the form
\be
H_I(t) = \int d^3x \sqrt{g}\sum_n \frac{V^{(n)} \phi^n}{n!}\,,
\ee
the Feynman rules are well known. In the context of cosmology where the metric depends on time, it is useful to have either a full real space representation or a mixed representation where we use momentum space for the spatial direction but keep the time dependence explicit. (For more details on how to find the Feynman rules in a cosmological context see for example \cite{vanderMeulen:2007ah} or the appendix of \cite{Leblond:2010yq}). 
We will use the Keldysh basis for the propagators
\bea
F(x,y) &  = & -\frac{i}{2}\left(G^{-+}(x,y) + G^{+-}(x,y)\right) \nn\; ,\\
G^R(x,y) & = & \theta(x_0-y_0) \left(G^{-+}(x,y) - G^{+-}(x,y)\right)\nn\; ,\\
G^A(x,y) & = & \theta(y_0-x_0) \left(G^{+-}(x,y) - G^{-+}(x,y)\right)\; ,
\eea
where $F$ stands for the Keldysh propagator and $G^{R,A}$ are the
retarded and the advanced propagators respectively. The Wightman functions are
$G^{-+}(x,y) = i \langle \phi(x) \, \phi(y) \rangle \,,  G^{+-}(x,y) = i \langle \phi(y) \, \phi(x) \rangle$.
 In our computation, we
will use exclusively the mixed representation where we Fourier
transform only the spatial part of the propagator. We show
the resulting Feynman rules in Fig.~(\ref{prop}). 

The propagators can be obtained from the mode functions 
\be
\phi(x,t) = \int \frac{d^dp}{(2\pi)^d}  \left(a_p e^{i\vec{p}\cdot \vec{x}}U_p(t) + a_p^\dagger e^{i\vec{p}\cdot \vec{x}} U_p^*(t)\right)\; .
\ee
We will work in the Poincar\'e patch of dS with metric 
\be
\label{dsmetric}
\exd s^2 = \frac{- \exd \tau^2 + \exd x_i \exd x^i}{\tau^2}
\ee
where we have set the Hubble constant equal to 1 and the conformal time $\tau$ runs from $-\infty$ to 0. The number  of space-time dimensions is $D = d+1$ . In the Bunch-Davies (BD) vacuum, the modes are  
 \cite{Krotov:2010ma}
\be\label{mode}
 U_k(\tau) =\frac{i e^{i\pi/4} \sqrt{\pi}  (-\tau)^{d/2} }{2} e^{-\pi\mu/2}H^{(1)}_{i\mu}(x) \equiv \frac{1}{\sqrt{2}} (-\tau)^{d/2} h(x,\mu) \,,
 \ee
where $x=-k\tau$,  $\mu  = \sqrt{ m^2 - (d/2)^2}$, we have dropped the irrelevant phase and 
\be
h(x,\mu) = \sqrt{\frac{\pi}{2}}e^{-\pi\mu/2}H^{(1)}_{i\mu}(x) \;.
\ee
We work in the principal series where $\mu$ is real and the order of the Hankel function is imaginary. 
The function $h$ is symmetric in the mass $h(x,\mu) = h(x,-\mu)$
and has the following asymptotic behavior for large and small values of $x$
\be\label{expansion}
h(x,\mu) \sim \left\{\begin{array}{cc} \sqrt{\frac{1}{x}} e^{i(x
 - \frac{\pi}{4})} &\;\;\;\; ; x\gg 1\\
 A_+ x^{i\mu} + A_- x^{-i\mu} &\;\;\;\; ; x\ll1 
\end{array}\right. 
\ee
where
$A_\pm$ are specific functions satisfying the following useful identities
\begin{align}\label{C}
|A_+|^2 - |A_-|^2& = 
\frac{1}{\mu}\,, &
 |A_+|^2 + |A_-|^2 & = 
\frac{1}{\mu} \coth(\pi\mu) \,.
\end{align}
In terms of those modes, the propagators are
 \bea\label{FG}
F(k,\tau_1,\tau_2) & = & (\tau_1\tau_2)^{d/2} f(k\tau_1,k\tau_2)  = \frac12 (\tau_1\tau_2)^{d/2}\ \Re \left( h(-k\tau_1,\mu) h^*(-k\tau_2,\mu) \right)\,,\\
G^R(k,\tau_1,\tau_2) & = & \theta(\tau_1-\tau_2)\tau_1^{d/2}\tau_2^{d/2}{g^{R}}(k\tau_1,k\tau_2) =  -  (\tau_1\tau_2)^{d/2}
\theta(\tau_1-\tau_2)\ \Im \left( h(-k\tau_1,\mu) h^*(-k\tau_2,\mu)\right)\,.\nn
\eea
We will often factor out $(\tau_1\tau_2)^{d/2}$ as well as any time ordering function and  denote the rest of the propagator by non-capitalized letters $g^R$ and $f$. 
We can expand the propagator for small value of the arguments $-k\tau_{1,2}\ll 1$ using Eq.~(\ref{expansion}) to get
\bea\label{small}
g^{R}& \approx &  \frac{i}{2 \mu}\left( \left(\frac{\tau_1}{\tau_2}\right)^{i\mu} -  \left(\frac{\tau_1}{\tau_2}\right)^{-i\mu}\right)\,,\\
f&\approx & \frac{1}{4}\left( 2 A_+ A_-^* (k^2\tau_1\tau_2)^{i\mu} + 2 A_-A_+^*(k^2\tau_1\tau_2)^{-i\mu} + \frac{1}{\mu}  \coth(\pi\mu)\left( \left(\frac{\tau_1}{\tau_2}\right)^{i\mu} +  \left(\frac{\tau_1}{\tau_2}\right)^{-i\mu}\right)\right)\,.\nn
\eea
We see that for small $k\tau_{1,2}$ both propagators have $k$
independent oscillating terms. In the loop computation these
oscillating terms can interfere destructively. When this happens one
gets a secularly divergent log in time $\ln k\tau$. 
We will compute this log contribution to the 1-loop diagrams below and we will then proceed to resum the 1-loop bubble diagrams.

\section{The Cubic 1-loop Correction}
\label{explicit}

In this section we compute the 1-loop correction from a cubic interaction to 
the retarded and the Keldysh propagator.  

\subsection{The Retarded Propagator}

\begin{figure} [ht]
\begin{center}
\includegraphics[width=0.4\textwidth,angle=0]{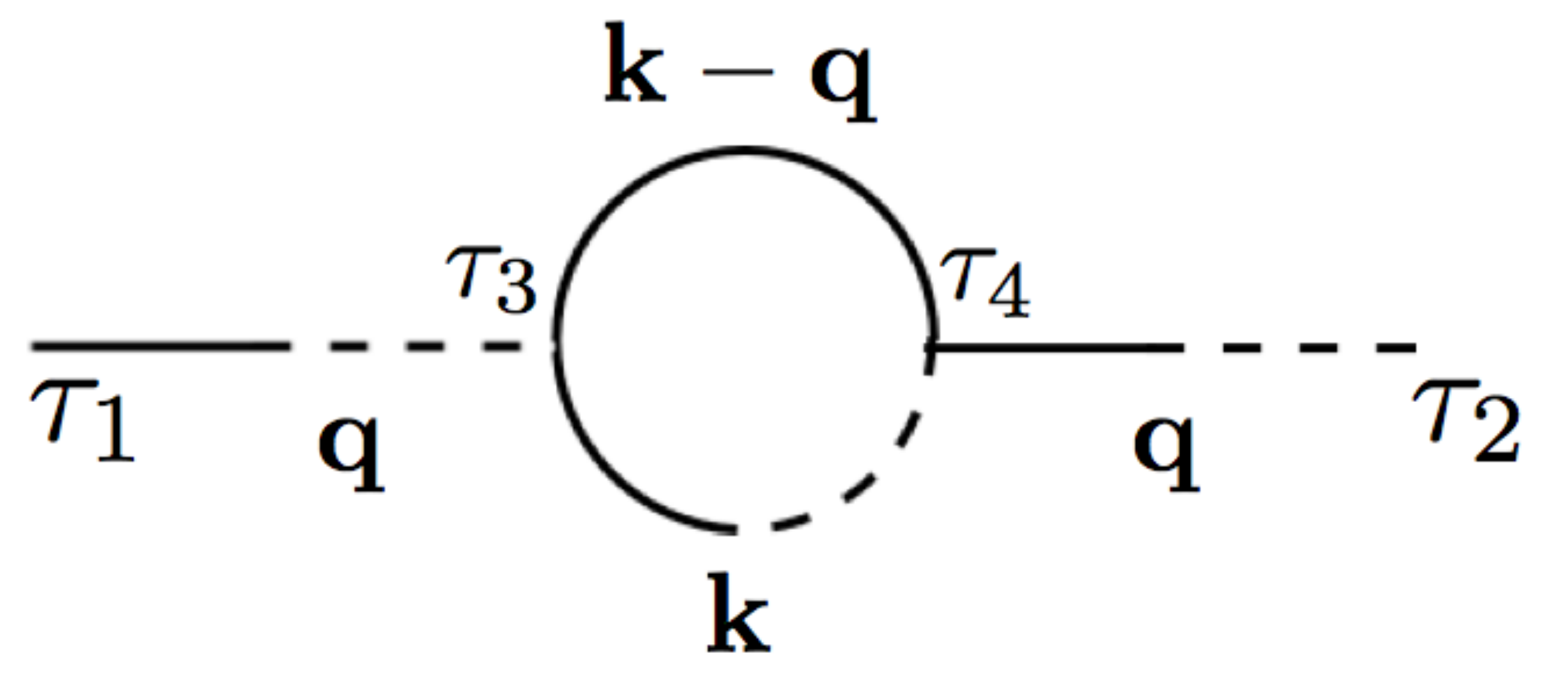}
\caption{The 1-loop $\phi^3$ correction to the retarded propagator; this is $-i G^{R1}(q,\tau_1,\tau_2)$.}\label{loopphi3}
\end{center}
\end{figure}

Let us start with the 1-loop correction to the retarded propagator from a cubic coupling $\frac{\lambda\phi^3}{3!}$. There is only one diagram which is shown in Fig.~(\ref{loopphi3}).
The 1-loop correction to the retarded propagator is 
\bea\label{basiceq}
G^{R1}(q,\tau_1,\tau_2)&=&\lambda^2\int \frac{d^dk}{(2\pi)^d} \int_{-\infty}^{0}d\tau_3\int_{-\infty}^{0} d\tau_4 {1\over( \tau_3\tau_4)^{d+1}}G^{R0}(q,\tau_1,\tau_3) \\
&& \hspace{1.7in} \times G^{R0}(k,\tau_3,\tau_4)F^0(|\vec{k}-\vec{q}|,\tau_3,\tau_4)
G^{R0}(q,\tau_4,\tau_2)\,,\nn
\eea
where $G^{R0}$ is the tree level propagator while $G^{R1}$ is the 1-loop part. A similar notation is used for $F$. There is no log coming from $k=0$ since the field is massive. Instead, a logarithm arises from the 
region where $k \gg q$. 
To see this, we first factor out the overall factor $(\tau_1\tau_2)^{d/2}$ from (\ref{basiceq}) 
to obtain
\bea
g^{R1} &\approx&
\lambda^2\int_q^{\infty}
\frac{d^dk}{(2\pi)^dk^d} \int_{k\tau_2}^{k\tau_1}dx_3
\int_{k\tau_2}^{x_3}dx_4(x_3x_4)^{d/2-1}g^{R0}(q\tau_1,{q\over
  k}x_3)g^{R0}(x_3,x_4)\nn \\
&&\hspace{2in} \times f^0(x_3,x_4)g^{R0}({q\over k}x_4,q\tau_2)\,.\nonumber
\eea
where we have defined $x_3 = k\tau_3$ and $x_4 =  k\tau_4$. 
We expand the two external propagators for 
 $-q\tau_{1,2} \ll1$ and $-\frac{q}{k}x_{3,4}\ll1$ as
 \bea
 g^{R0}(q\tau_1,{q\over k}x_3) &\sim &\frac{i}{2\mu} \left( \left(\frac{\tau_1}{x_3}\right)^{i\mu} k^{i\mu}-  \left(\frac{x_3}{\tau_1}\right)^{i\mu}k^{-i\mu}\right)\,,\nn\\
  g^{R0}({q\over k}x_4,q\tau_2) &\sim &\frac{i}{2\mu}\left( \left(\frac{x_4}{\tau_2}\right)^{i\mu} k^{-i\mu}-  \left(\frac{\tau_2}{x_4}\right)^{i\mu}k^{i\mu}\right)\,.
 \eea
In the product of the two $g^{R0}$ above, there are two terms 
which interfere destructively
\bea\label{inter}
g^{R1} &\sim& \lambda^2 \left(\frac{i}{2\mu}\right) ^2
\int_q^{\infty}
\frac{d^dk}{(2\pi)^dk^d} \int_{k\tau_2}^{k\tau_1}dx_3
\int_{k\tau_2}^{x_3}dx_4(x_3x_4)^{d/2-1} g^{R0}(x_3,x_4)f^0(x_3,x_4)
\nn \\
&& \hspace{2in}\times \left(  \left(\frac{\tau_1}{\tau_2}\right)^{i\mu} \left(\frac{x_4}{x_3}\right)^{i\mu} +  \left(\frac{\tau_2}{\tau_1}\right)^{i\mu} \left(\frac{x_3}{x_4}\right)^{i\mu}\right)\,.
\eea
In the limit that $-k\tau_1\rightarrow 0$ and $-k\tau_2 \rightarrow \infty $, the time integral is independent of $k$
\be
\int_{k\tau_2}^{k\tau_1}dx_3 \int_{k\tau_2}^{x_3}dx_4 \sim \int_{-\infty}^{0}dx_3 \int_{-\infty}^{x_3}dx_4 = \hf \int_{-\infty}^{0}dx_3\int_{-\infty}^{0}dx_4\,.
\ee
Note that this is only valid for $-k\tau_1\ll1 $ and $-k\tau_2 \gg1$ which imposes new limits on the $k$ integral.
Since $g^{R0}$ is odd under the exchange of two arguments while $f^0$ is even, the two terms are the same up to a minus sign. Hence
\be
g^{R1} \sim \frac{i\sigma}{2 \mu} \left(  \left(\frac{\tau_1}{\tau_2}\right)^{i\mu} -  \left(\frac{\tau_2}{\tau_1}\right)^{i\mu}\right) \int_{-1/\tau_2}^{-1/\tau_1} \frac{dk }{k}\,,
\ee
with
\bea\label{sigma}
\sigma& \equiv  &\frac{i \lambda^2 J^{gf}(\mu) S_{d-1}}{(2\pi)^d 2 \mu}\,,\nn\\
J^{gf}(\mu) & \equiv & \hf \int_{-\infty}^{0}dx_4 \int_{-\infty}^{0}dx_3 x_3^{-i\mu} x_4^{i\mu} (x_3x_4)^{d/2-1} g^{R0}(x_3,x_4)f^0(x_3,x_4)\,,
\eea
where $S_{d-1}$ is the volume 
 of the $d-1$ dimensional sphere. As promised this part of the diagrams is UV convergent and the momenta integral only give rises to a logarithm of ratios between the two times. 
 
Altogether, the retarded propagator at 1-loop in the limit $\tau_1 > \tau_2$ and for $-q\tau_{1,2} \ll 1$ goes like
\be\label{1loop}
G^{R(0+1)}(q,\tau_1,\tau_2)  \approx G^{R0}(q,\tau_1,\tau_2) \left[1- \sigma \ln \frac{\tau_1}{\tau_2} \right]\; .\ee
Some comments are in order regarding the (multiple) approximations done in obtaining this simple answer. We evaluated this loop contribution assuming $\tau_1 \gg \tau_2$ and we took the external momenta to be small with $q\ll -1/\tau_1$. We neglected the $k< q$ part of the integral as well as approximating the limits in the $J(\mu)$ integral. Both approximations will lead to extra $k$ dependence which then lead to power law contribution of the type $ \mathcal{O}(\tau_1/\tau_2)$ (which are negligible in the late $\tau_1\rightarrow 0$ limit). 
Using Eq.~(\ref{FG}), 
\bea\label{gf}
J^{gf}(\mu) & = & \frac{i }{16}\int _0^\infty dx_4dx_3 x_3^{-i\mu} x_4^{i\mu} (x_3x_4)^{d/2-1} (h(x_3) h^*(x_4) - h(x_3)^*h(x_4) ) \nn \\
&& \hspace{1cm}\times (h(x_3) h^*(x_4) + h(x_3)^*h(x_4) )\nn\\
&=&-\frac{i}{16} \left( |g(\mu)|^2 - |g(-\mu)|^2\right)\,,\nn\\
& =& \frac{i}{16} |g(-\mu)|^2 (1-e^{-2\pi\mu})\,,
\eea
with
\be\label{gmu}
g(\mu) = \int_0^\infty dx  x^{d/2-1+i\mu} h(x,\mu)^2\,.
\ee
This is the same formula as in \cite{Krotov:2010ma}.
The last line was obtained using $|g(\mu)|^2 = e^{-2\pi\mu} |g(-\mu)|^2$
 which follows directly from the properties of Hankel functions.
The 1-loop coefficient is then given by 
\be
\sigma = - \frac{\lambda^2 S_{d-1}}{32\mu (2\pi)^d} |g(-\mu)|^2 (1-e^{-2\pi\mu})\,,
\ee
which is real and negative.  In section \S~\ref{decay}, we will analyze this integral further. 
\subsection{The Keldysh Propagator}
\label{Keldysh}
\begin{figure} [ht]
\begin{center}
\includegraphics[width=0.4\textwidth,angle=0]{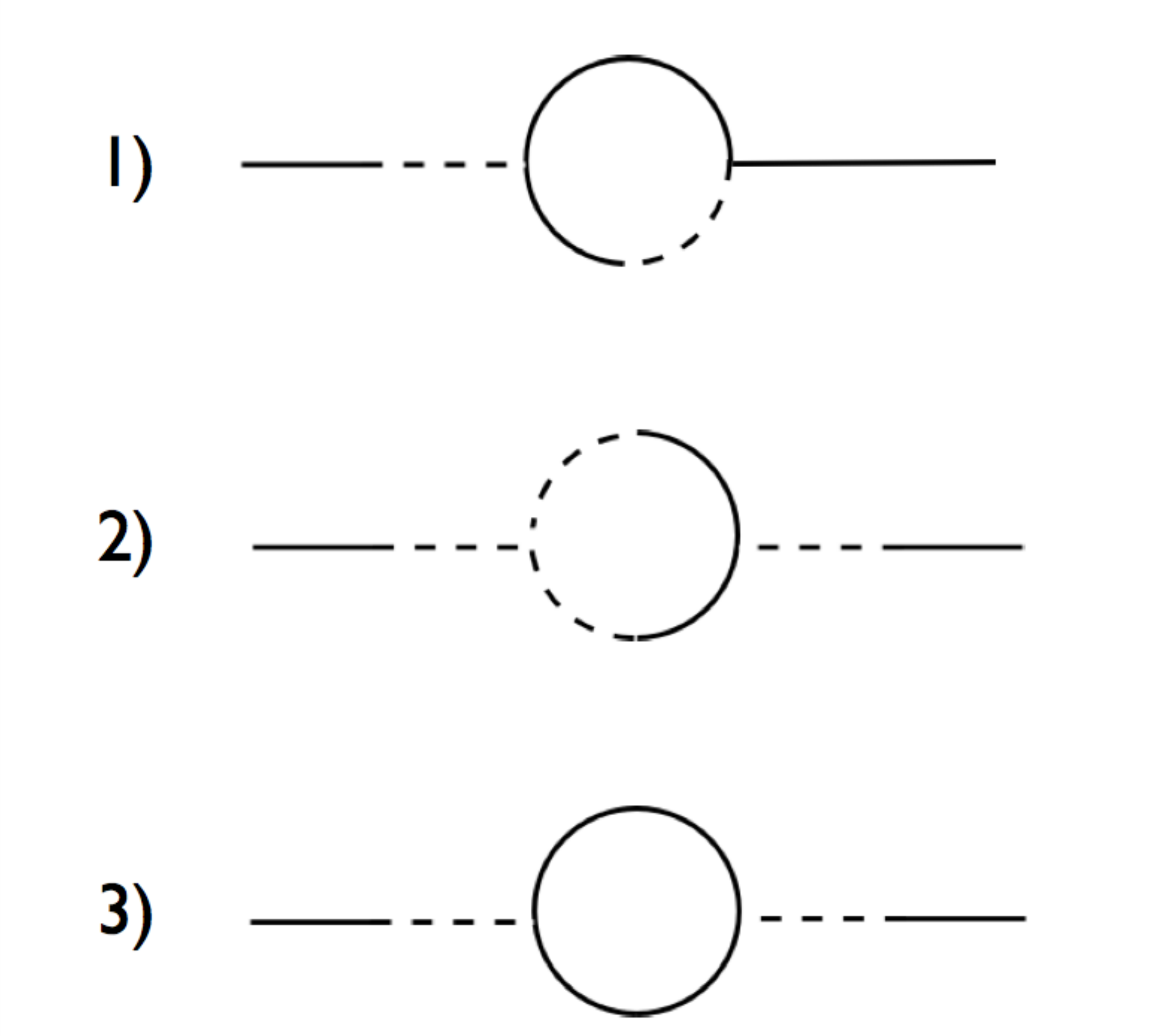}
\caption{
Corrections to the F propagator. The time reversal (reflection) of all diagrams needs to be included as well.}\label{loopF}
\end{center}
\end{figure}
There are three 
1-loop graphs contributing to the Keldysh
propagator as shown in Fig.~(\ref{loopF}) (there are five diagrams including time reversals). 
These diagrams may be computed using the same  approximations and methods as in the last section. 

For all diagrams, one obtains equations similar to
Eq.~(\ref{inter}) with the following differences. First, we have either two
$g^R$, two $f$ or one $g^R$ and one $f$ in the time integral. We denote the new integrals as
$J^{gg}$ and $J^{ff}$ respectively. Second, the limits on the time
integrals are different. 
For diagrams (1), the limits of integration are
\be
\int_{-\infty}^{k\tau_1} dx_3\int _{-\infty}^{x_3} dx_4\,.
\ee
and in the time reversal, we replace $\tau_1$ by $\tau_2$. For diagram (2), the limits are now
\be
\int_{-\infty}^{x_4} dx_3\int _{-\infty}^{k\tau_2} dx_4\,.
\ee
and this is unchanged for the time reversed diagram (we took $\tau_1 > \tau_2$). Finally for diagram (3), the limits of integration are
 \be
\int_{-\infty}^{k\tau_1} dx_3\int _{-\infty}^{k\tau_2} dx_4\,.
\ee
and there is no time reversal. 
Demanding that these integral be $k$-independent require both $-k\tau_{1,2}\ll1$. This gives an upper limit in the momenta $k$ but the lower limit remains at $q$ (to be contrasted to what happen for the retarded propagator).  

The Keldysh propagator in the small $-q\tau\ll1$ is given in Eq.~(\ref{small}); here we further neglect the $q\tau$ pieces compared to the other terms
\be\label{qtzero}
f^0(q,\tau_1,\tau_2)= \frac{\coth\pi\mu}{4\mu} \left( \left(\frac{\tau_1}{\tau_2}\right)^{i\mu} +  \left(\frac{\tau_1}{\tau_2}\right)^{-i\mu}\right)\,.
\ee
Approximating $f^0$ to this term, one finds that the 1-loop correction from diagram (2) and (3) is 
\be
f^{1} =   -f^0(q,\tau_1,\tau_2) (\sigma_2 + \sigma_3)  \ln -q\tau_{2}\,,
\ee
where again $\tau_1>\tau_2$ and
\begin{align}
\sigma_2 & =  \frac{\lambda^2 S_{d-1}}{4\mu \coth(\pi\mu) (2\pi)^d} J^{gg}(\mu)\,, &
\sigma_3 & = - \frac{ \lambda^2  S_{d-1}}{\mu \coth(\pi\mu) (2\pi)^d}  J^{ff}(\mu)\,.
\end{align}
We would like to emphasize the relative sign between $\sigma_2$ and $\sigma_3$ due to the presence of 4 retarded propagators (each with their $-i$ factors) as opposed to 2 and also note the factor of $1/4$ from the vertex. Also we have included the time reversed diagram for $\sigma_2$ which amounts to an extra factor of 2 and there is a symmetry factor of $1/2$ for both diagrams. 
 Now both $\sigma_2$ and $\sigma_3$  are real and they lead to an imaginary
 shift to the mass. It is worth pointing out that the sum of the two terms does
 not contain a divergence of the type $\int h(x) h^*(x)$ due to the identity
\be\label{ff}
4 J^{ff} - J^{gg} =\frac{1}{4}\left( |g(\mu)|^2 +|g(-\mu)|^2\right)\,.
\ee
Finally, diagram 1 in Fig.~(\ref{loopF}) also has a very similar structure. We expand the Keldysh propagator in the last leg as before and the final result is 
\be
f^{(0+1)} (q,\tau_1,\tau_2)  =  f^0(q,\tau_1,\tau_2)\left[1-  \sigma_1(\ln (-q\tau_1) + \ln (-q\tau_2))\right]\,,
\ee
with
\bea\label{sigma1}
\sigma_1 & = & \frac{i \lambda^2 J^{gf}(\mu) S_{d-1}}{(2\pi)^d 2 \mu}\,,
\eea
Adding all the contributions together and writing $\ln (-q\tau_1) = \ln \frac{\tau_1}{\tau_2} + 2\ln (-q\tau_2)$, we get 
\be
f^{(0+1)} (q,\tau_1,\tau_2)  =  f^0(q,\tau_1,\tau_2)\left[1-  \sigma_1\ln \frac{\tau_1}{\tau_2} - ( 2\sigma_1 +\sigma_2+\sigma_3) \ln (-q\tau_2))\right]\,,
\ee
Using the properties of the integrals $J^{gg}(\mu)$, $J^{ff}(\mu)$ and $J^{gf}(\mu)$ in Eq.~(\ref{gf}) and Eq.~(\ref{ff}), one can show that 
\be
2\sigma_1 +\sigma_2+\sigma_3 = 0\; .
\ee
Therefore the full answer is just 
\be
f^{(0+1)} (q,\tau_1,\tau_2)  =  f^0(q,\tau_1,\tau_2)\left(1-  \sigma_1\ln \frac{\tau_1}{\tau_2} \right)
\ee
and $\sigma_1$ is the imaginary shift to the mass.  It is exactly equal to the shift we found for the retarded propagator in Eq.~(\ref{sigma}).  
 
\section{Resummation of Large Logarithms}\label{resum}
In the last section we computed the 1-loop contribution to the
retarded and to the Keldysh propagators in $\lambda\phi^3$ theory.  We found that the
result contains a potentially large logarithm that is either a ratio of two times or a ratio of the latest time to the external momenta. In order to obtain this result
we had to assume that $\tau_1\gg \tau_2$ or that $q\tau_{1,2}\ll1 $ and the logarithmic
contributions are then dominant and large.  These large logs cause trouble as they naively
signal a  breakdown of perturbation theory
at late time. To address this problem, we will now resum these 1-loop
logarithms by using two different methods, both of which give the same
result which is finite and well behaved at late times $\tau_1\rightarrow 0^-$.

The resummation technique demonstrates that the 1-loop contribution resums
to a time independent (in this case imaginary) shift to the mass.
Indeed, if one performs this computation by Wick rotating to the sphere as in \cite{Marolf:2010zp}, there is no logarithmic time dependence and the 1-loop simply gives a finite contribution to the self-energy which shifts the mass. So the results of our resummation shows clearly that the same answer is obtained using the in-in formalism in the Poincar\'e patch versus using the sphere. On the other hand, our resummation methods can easily be extended to non-equilibrium situations that may not be captured by a computation on the sphere; for example global de Sitter space
 \cite{Krotov:2010ma} or  an unstable vacuum (as opposed to the Bunch-Davies vacuum). 

In the case of the retarded propagator we are in a fortunate
situation because there is only one diagram at 1-loop and therefore it is a
simple matter to resum.  Large infrared logarithms have been argued to be well captured by the classical approximation 
(see for example \cite{vanderMeulen:2007ah}) or by semi-classical
methods such as the stochastic approach \cite{Starobinsky:1994bd} (see also \cite{Woodard:2005cw,   Prokopec:2007ak, Miao:2006pn}). 
Here are two different ways to resum the
1-loop correction to the retarded propagator where the errors are easily quantifiable and the next order correction is in principle calculable. 

\begin{enumerate}
\item In the first method we will set-up a
Schwinger-Dyson like non-perturbative equation and solve it by making
an ansatz for the 1-loop exact retarded propagator.  

\item 
In the second method we explicitly
compute the $n-$loop bubble diagrams and we resum them all (this is for the retarded propagator). 
\end{enumerate}

It would also be interesting to use dynamical renormalization group methods (see \cite{Chen:1995ena, Boyanovsky:1998aa, Boyanovsky:2003ui} and \cite{Boyanovsky:2004gq, Burgess:2009bs, Burgess:2010dd} for some applications in cosmology) to perform the resummation but we leave this to further work.

\subsection{Schwinger-Dyson Equation for the Retarded Propagator}

\begin{figure} [ht]
\begin{center}
\includegraphics[width=1\textwidth,angle=0]{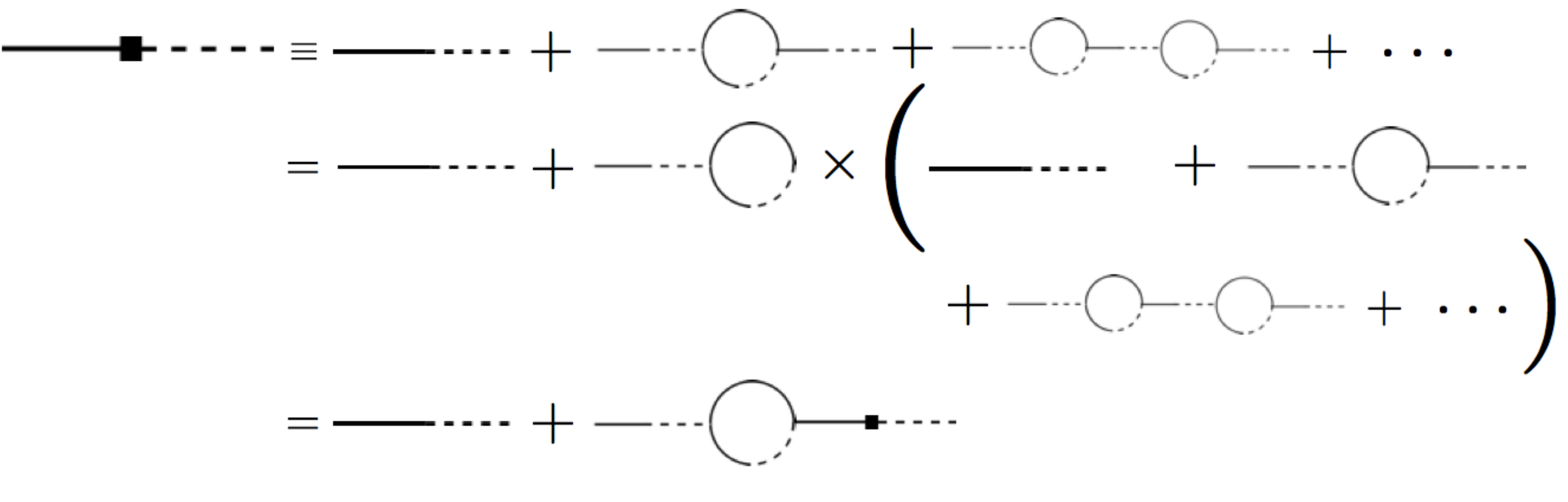}
\caption{The Schwinger-Dyson equation for the retarded propagator. The line with the square is meant to represent the fully resummed propagator. }\label{SD}
\end{center}
\end{figure}

Considering only the bubble diagram, it is simple to set-up a non-perturbative equation for the fully resummed propagator a la Schwinger-Dyson (see Fig.~\ref{SD}).
The resulting equation is of the form Eq.~(\ref{basiceq}) with the last propagator replaced by the full resummed one. 
We make the ansatz that the
mode functions, which were originally written as Eq.~(\ref{mode}), should become
\be
 U_{k}(\tau) \rightarrow (-\tau)^{d/2} h^R(x,\mu)\,,
 \ee
where $h^R(k\tau)$ are some new functions.
Correspondingly, the propagator is
now of the form
\bea
G^{R}(q,\tau_1,\tau_3)=\theta(\tau_1-\tau_3)(\tau_1\tau_3)^{d/2}{g^{R}}(q\tau_1,q\tau_3)\,.
\eea
We can write the Schwinger-Dyson equation as 
\bea
&&g^{R}(q\tau_1,q\tau_2)= g^{R0}(q\tau_1,q\tau_2)\nonumber
\\
&&+\lambda^2\int_q^{\infty} \frac{d^dk}{(2\pi)^d} \int_{\tau_2}^{\tau_1}d\tau_3 \int_{\tau_3}^{\tau_2}d\tau_4(\tau_3\tau_4)^{d/2-1}g^{R0}(q\tau_1,q\tau_3)g^{R0}(k\tau_3,k\tau_4)f^0(k\tau_3,k\tau_4)
g^{R}(q\tau_4,q\tau_2)\,,\nonumber
\eea
where as before we have taken the limit $k\gg q$. 
We
perform the Fourier expansion of
${g^{R}}(q\tau_4,q\tau_2)$ in terms of $q\tau_4$, and in particular, we consider a Fourier component which is of the form $(q\tau_4)^{ia}h_a(q\tau_2)$. 
The modes $a=\pm\mu$ received a large  enhancement from the integral proportional to $\ln\tau_1/\tau_2$ as we found in Eq.~(\ref{1loop}).
Now {\it ex hypothesi}, the resummed propagator should have a good perturbative
expansion for all time, and hence we conclude that there are no
Fourier components of
${g^{R}}(q\tau_4,q\tau_2)$ proportional to $(q\tau_4)^{\pm i\mu}$. 
Since at leading order, such terms are indeed
present,  what must therefore happen is that the exponent of these Fourier components is corrected by a shift which is proportional to the coupling. 

The mode functions which receive the largest corrections are thus the ones proportional to
$\phi(k,\tau)\sim (k\tau)^{\pm i\mu}$; we assume that these mode functions are corrected to
be of the form $\phi(k,\tau)\sim(k\tau)^{-b \pm i\mu}$ for some $b$. 
These will produce large  corrections to the propagator which depend on the integrals
 \bea
I_{\pm}& =&\int_{-1/\tau_2}^{-1/\tau_1} {dk\over k^{1-b}} \int_{-\infty}^{0}dx_3 \int_{-\infty}^{0}dx_4 {(x_3 x_4)^{d/2-1}}x_3^{\mp i\mu}x_4^{\pm i\mu}{g^{R0}}(x_3,x_4){f^{0}}(x_3,x_4) x_4^{-b}
\nonumber\\
&\equiv &{1\over b}\left(\left(\frac{-1}{\tau_1}\right)^{b}-\left(\frac{-1}{\tau_2}\right)^{b} \right)J^{gf}(\pm\mu, b)\,,
\eea
where $J^{gf}(\pm\mu, b)$ is the integral defined in Eq.~(\ref{sigma}) with the extra $x_4^{-b}$ in the integrand. We have employed the same approximations as in section \S~\ref{explicit}
and $\tau_1\gg \tau_2$. We split the resummed propagator into the resonantly enhanced and the resonantly unenhanced modes by
\bea\label{ansatz}
{g^{R}}(q\tau_4,q\tau_2)
=\frac{i}{2\mu} \left(  \left(\frac{\tau_4}{\tau_2}\right)^{i\mu} -  \left(\frac{\tau_4}{\tau_2}\right)^{-i\mu}\right) \left(\frac{\tau_4}{\tau_2}\right)^{-b}
+{\gamma^{R}}(q\tau_4,q\tau_2)
\eea
and ${\gamma^{R}}$ contains no terms with Fourier component $(q\tau_4)^{\pm i\mu}$.
The terms in ${\gamma^{R}}$ does not lead to large integrals on the
right hand side, and can be dropped at leading order on the RHS.
Since every other term
in the equation only contains the Fourier components $\tau_1^{\pm i\mu}$,
it follows that ${\gamma^{R}}$ is zero at leading order. With this ansatz for the full retarded propagator, the Schwinger-Dyson equation becomes  
\be
g^R  = g^{R0} + g^{R0}  \frac{\sigma(b)}{b} \left(\left(\frac{\tau_2}{\tau_1}\right)^{b}-1 \right)\,,
\ee
which of course exactly reproduces the 1-loop answer Eq.~(\ref{1loop}) when $b = 0$. The factor $\sigma(b)$ is as given in Eq.~(\ref{sigma}) with again the extra factor of $x_4^{-b}$ in the integrand
\be\label{sigmab}
\sigma(b)  \equiv  \frac{i \lambda^2 S_{d-1}}{(2\pi)^d 4 \mu} \int_{-\infty}^{0}dx_4 \int_{-\infty}^{0}dx_3 x_3^{-i\mu} x_4^{i\mu} (x_3x_4)^{d/2-1} g^{R0}(x_3,x_4)f^0(x_3,x_4) x_4^{-b}\,.
\ee
 Now we see that we can consistently solve the SD equation by  setting
\be
b = \sigma(b)
\ee
Which in general is some transcendental equation to be solved for $b$.  We can approximate $\sigma(b) \approx \sigma$ (given by Eq.~(\ref{sigma})). This amount to resumming only the bubble diagrams and nothing else. 
The upshot of this computation is that the ansatz
Eq.~(\ref{ansatz}) consistently 
solves  the SD equation for $b= \sigma(b)$.  This modification of the
propagator can be seen as a modification of the modes.  As a result, the same
ansatz can be used for getting the resummed $F$  propagator.
We will rederive this
result in subsection \S~\ref{direct}, by doing direct resummation of 1-loop chain
diagrams for the retarded propagator.  

\subsection{The Keldysh Propagator}
\begin{figure} [ht]
\begin{center}
\includegraphics[width=0.8\textwidth,angle=0]{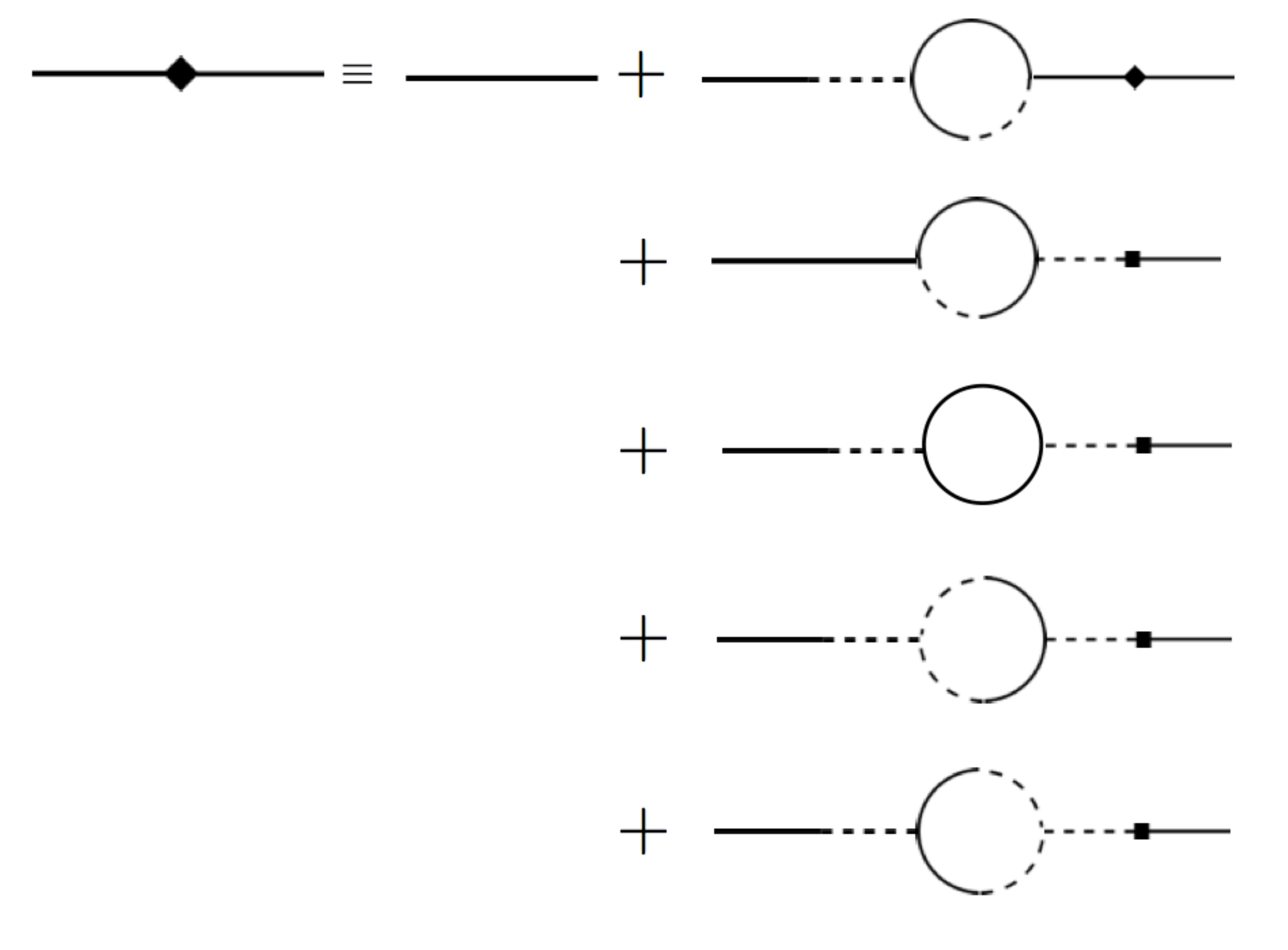}
\caption{The Schwinger-Dyson equation for the Keldysh propagator}\label{SDkeldysh}
\end{center}
\end{figure}

We now turn to the $F$ propagator. 
Unlike the retarded propagator, we have
more bubbles to account for in the Schwinger-Dyson equation, see Fig.~(\ref{SDkeldysh}).
The structure is completely similar as in the $G^R$ case. For example, including only the first diagram, the SD equation reads
\bea
&&f(q\tau_1,q\tau_2)= f^{0}(q\tau_1,q\tau_2)\nonumber
\\
&&+\lambda^2\int_q^{\infty} \frac{d^dk}{(2\pi)^d} \int_{-\infty}^{\tau_1}d\tau_3 \int_{-\infty}^{\tau_3}d\tau_4(\tau_3\tau_4)^{d/2-1}g^{R0}(q\tau_1,q\tau_3)g^{R0}(k\tau_3,k\tau_4)f^0(k\tau_3,k\tau_4)
f(q\tau_4,q\tau_2)+\cdots\,.\nonumber
\eea
We can consistently solve the SD equation by starting with the same ansatz as we had for $g^R$
\bea\label{ansatz2}
{f}(q\tau_4,q\tau_2)
=\frac{\coth(\pi\mu)}{4\mu} \left(  \left(\frac{\tau_4}{\tau_2}\right)^{i\mu}  + \left(\frac{\tau_4}{\tau_2}\right)^{-i\mu}\right) \left(\frac{\tau_4}{\tau_2}\right)^{-b}
\eea
for some undetermined $b$. There is five 1-loop computations to perform and we denote them 1 to 5 going from top to bottom in Fig.~(\ref{SDkeldysh})
\bea
f_1 &=& \sigma_1(b) \frac{f^0}{b}  \left( \left(\frac{\tau_2}{\tau_1}\right)^b - (-q\tau_2)^b\right)\; ,\\
f_2 &=& \sigma_1(b) \frac{f^0}{b}  (1 -(-q\tau_2)^b)\; ,\nn\\
f_3 &=& \sigma_3(b) \frac{f^0}{b}  (1 -(-q\tau_2)^b)\; ,\nn\\
f_4 &=& \frac{\sigma_2(b)}{2} \frac{f^0}{b}  (1 -(-q\tau_2)^b)\; ,\nn\\
f_5 &=& \frac{\sigma_2(b)}{2} \frac{f^0}{b}  (1 -(-q\tau_2)^b)\; .\nn
\eea
Each $\sigma_i(b)$ are exactly as given in section (\ref{Keldysh}) but with an extra factor of $x_4^{-b}$ in the integrand. 
The SD equation then becomes
\be
f  =f^0 + f^0 \frac{\sigma_1(b)}{b} \left(\left(\frac{\tau_2}{\tau_1}\right)^{b}- 1\right)  + \frac{f^0}{b}\left(2\sigma_1(b) +\sigma_2(b)+\sigma_3(b)\right) \left( 1 - (-q\tau_2)^b\right)   
\ee
where the first term is the tree level contribution and we have rearranged the contribution from the first and second diagram. As before one can show that for any $b$
\be
2\sigma_1(b) +\sigma_2(b)+\sigma_3(b) = 0\; .
\ee
The SD equation can consistently be solved by setting
\be
b = \sigma_1(b)\; .
\ee 
Neglecting all higher order terms and keeping only bubble diagrams we have that $\sigma_1 = b$. Note that the shift to $F$ is the same as the shift to $G^R$, that is $\sigma_1 = \sigma$.  

\subsection{Direct Resummation}\label{direct}
A direct computation of the 2-loop chain diagrams for the retarded propagator is simple to do, one simply replaces the last propagator by the 1-loop result
\bea
g^{R2} &=& \lambda^2\int_q^{\infty} \frac{d^dk}{(2\pi)^dk^d}
\int_{k\tau_2}^{k\tau_1}dx_3
\int_{k\tau_2}^{x_3}dx_4(x_3x_4)^{d/2-1}g^{R0}(q\tau_1,{q\over
  k}x_3)g^{R0}(x_3,x_4)\nn\\
&&\hspace{2in}\times f^0(x_3,x_4)g^{R1}({q\over k}x_4,q\tau_2)\,.
\eea
The 1-loop propagator contains a $\ln k\tau_2$ which when integrated over $k$ will give a $(\ln \frac{\tau_1}{\tau_2})^2$. In the large time difference limit $ |\tau_1/\tau_2| \ll1$, the log term will dominate over other corrections (it is not enough to simply demand $-q\tau_{1,2} \ll1$).  Finally we approximate $\ln \frac{x_4}{k\tau_2} \sim - \ln (-k\tau_2) $. The $\ln x_4$ is dropped because it contributes a subleading $\ln\tau_1/\tau_2$ in the final answer.
Hence
\bea
g^{R2} & = & \frac{i}{2\mu} \sigma^2 \left(  \left(\frac{\tau_1}{\tau_2}\right)^{i\mu} -  \left(\frac{\tau_2}{\tau_1}\right)^{i\mu}\right) \int_{-1/\tau_2}^{-1/\tau_1} \frac{d^dk}{(2\pi)^dk^d} \ln (-k\tau_2)\nn\\
 & = & g^{R0} \sigma^2 \frac{\left(\ln \frac{\tau_2}{\tau_1}\right)^2}{2}\,.
\eea
A clear pattern emerges and the $n-$loop chain (bubble) diagrams is just 
 \be
 g^{Rn}  =  g^{R0} \sigma^n \frac{\left(\ln \frac{\tau_2}{\tau_1}\right)^n}{n!}\,,
 \ee
and the sum exponentiates to give
\bea\label{resummed2}
g^R &= &g^{R0} e^{-\sigma \ln \tau_1/\tau_2}\nn \\
 &= &\frac{i}{2\mu} \left[\left(\frac{\tau_1}{\tau_2}\right)^{i\mu -\sigma} - \left(\frac{\tau_1}{\tau_2}\right)^{-i\mu -\sigma}\right]\,.
\eea
The direct resummation of the Keldysh propagator does not work as well partly because of the $q\tau$ terms in the propagator. This is not really that surprising as it often happens that one can consistently solve the SD equations (as we did for the Keldysh propagator) but it still proves impossible to perturbatively resum diagrams.

\section{The Decay Rate}
\label{decay}
The resummed propagator (for e.g.~the retarded propagator Eq.~(\ref{resummed2})) has exactly the same form as the bare one. Using mode functions representation from Eq.~(\ref{FG})
\be
g^R(k,\tau_1,\tau_2) =  \Im \left( h(-k\tau_1,\mu) h^*(-k\tau_2,\mu)\right)\, =  \frac{\pi e^{-\pi\mu}}{2} \Im \left( H^{(1)}_{i\mu-\sigma}(x)H^{(2)}_{(i\mu-\sigma)^*}(x)\right)
\ee
we see that the resummation of bubble diagrams has lead to an imaginary shift to the mass.
\be
\mu \rightarrow \mu + i\sigma\,,
\ee
with
\be
\sigma = -  \frac{\lambda^2 S_{d-1}}{32\mu (2\pi)^d} |g(-\mu)|^2 (1-e^{-2\pi\mu}) \equiv -\frac{\lambda^2 S_{d-1}}{(2\pi)^d} K(\mu)\,.
\ee
Note the complex conjugation of the order parameter in the second Hankel function; one should not simply shift the mass in the propagator since there is complex conjugation going behind the scene. 
This formula is valid for any $\mu$ real and greater than 0.

\begin{figure} [ht]
\begin{center}
\includegraphics[width=0.6\textwidth,angle=0]{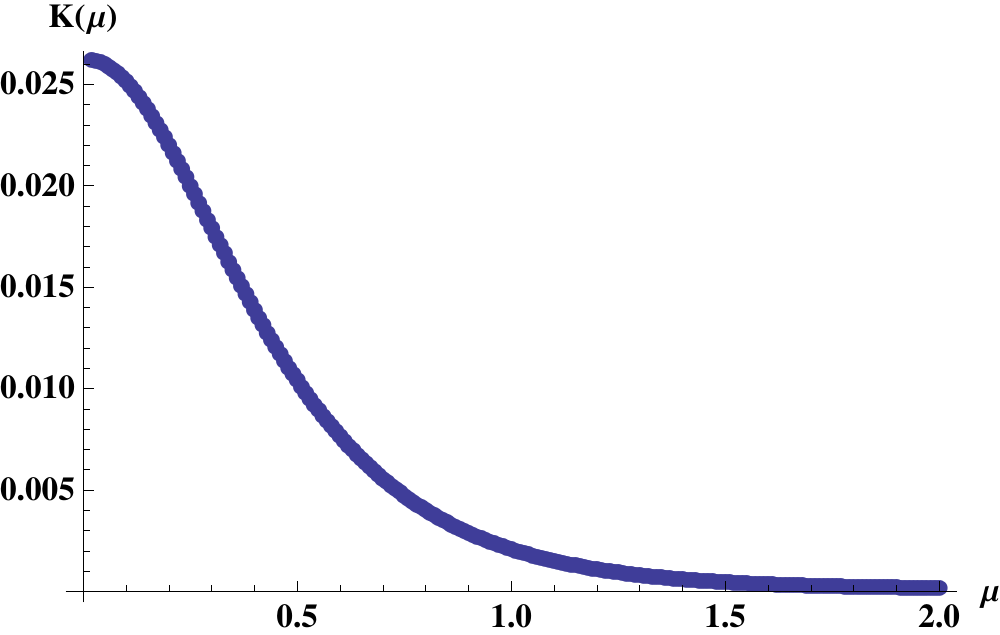}
\caption{The integral $K(\mu)$ as a function of $\mu$ for $d=3$.}\label{kmu}
\end{center}
\end{figure}

If $\mu \approx m  \gg 0$, then the propagators approach their flat space equivalent and the imaginary shift to the mass leads to a Breit-Wigner resonance of the form 
\be
\frac{1}{p^2 - m^2 + i m \Gamma}\; .
\ee
Therefore, in analogy with flat space results, we define the decay rate to
be 
\be
\Gamma \equiv -2 \sigma
\ee 
which is positive (as it should be) since $\sigma$ is negative.
It is simple to perform the integral $g(\mu)$ given by Eq.~(\ref{gmu}) numerically. The integrand oscillates wildly for large $x$ and we regulate it with an exponential cutoff.
In Fig.~(\ref{kmu}), we plot the function $K(\mu)$ as a function of $\mu$.
  We find that the decay rate is Boltzman suppressed as found in \cite{Bros:2008sq}.
\be
K(\mu)\xrightarrow[\mu \rightarrow \infty]{} e^{-\pi\mu} \; .
\ee
We find that the integral converge in any dimensions. For $\mu = 0$ or $m = \frac{dH}{2}$, the decay rate is finite and of order
\be
\Gamma \approx 0.05 \frac{\lambda^2 S_{d-1}}{(2\pi)^d}\; .
\ee
The angular factor of $S_{d-1}/(2\pi)^d$ is as usual for loop computations.

\section{Conclusion}

We have shown that there is indeed an imaginary shift to
the mass of a massive field with a $\phi^3$ interaction, as argued previously
by Marolf and Morrison \cite{Marolf:2010zp}.
In the in-in Poincar\'e calculation, the loop contribution receives an imaginary contribution that diverges at late time.  We have shown that one can consistently solve the Schwinger-Dyson equation by shifting the mass by an imaginary component.  This effectively resums all the bubble diagrams directly in the Lorentzian signature spacetime and  reproduces the 1PI resummation from the sphere.  

Many other resummation techniques could be used and it is worthwhile to explore them. From the Schwinger-Dyson equations, one can set-up a Boltzman kinetic equation. Alternatively, 
stochastic techniques \cite{Starobinsky:1994bd} or dynamical renormalization 
group techniques \cite{Boyanovsky:2004gq,Burgess:2009bs} could also prove 
to be very useful. 

Our results give additional support to the statement that interacting massive fields with 
$m> \frac{d H}{2}$ in de Sitter space are fundamentally unstable to decay. 
The reader may worry about the interpretation of this imaginary shift to the mass as a decay rate given that 
in de Sitter, there is no well defined notion of particles and the S-matrix is an ill defined quantity (the Jost vacua of \cite{Polyakov:2007mm} -- see also \cite{Akhmedov:2009ta}-- is the exception to this statement). In fact there is no contradiction as a decay rate is not an in/out quantity. One can never take a fundamentally unstable particle to $t\rightarrow -\infty$ to prepare an initial state. An unstable particle is not really a particle, it is a resonance. If one take the large mass limit of the resummed propagator in dS one recovers the flat space answer
\be
\frac{1}{p^2 - m^2 + i m \Gamma}
\ee
and clearly our imaginary shift to the mass then lead to the usual Breit-Wigner resonance.

Our notion of decay rate differs from the one used in \cite{Boyanovsky:2000hs,Boyanovsky:2004gq,Boyanovsky:2004ph} where they are looking for time dependence in correlation functions (in particular, they are looking for decay of the vacua to some other state). The case of conformally coupled scalar with cubic interactions was recently examined in \cite{Boyanovsky:2011xn}. In that paper, they both found a time dependent anomalous dimensions and a decay rate. For the quartic potential, they find no such decay rate. Again to contrast, we have examined the case of a massive field in the principal series which is perturbatively stable for cubic or quartic interactions but both cases show an imaginary contribution to the self-energy which we interpret as a decay rate. 

The case of global de Sitter is very interesting. 
Krotov and Polyakov also computed the 1-loop diagram in global de Sitter. They argued that the euclidean theory does not correctly capture the physics in this case.  As global de Sitter can roughly be viewed as two copies of Poincar\'e, the result for the 1-loop is very similar
\be\label{global}
F(k,\tau) \rightarrow F(k,\tau)^0\left( 1+ \sigma \ln \frac{\tau}{\tilde\tau_0}\right) 
\ee
where $\tilde\tau_0$ is the time at which one turns on 
the interaction in the lower (contracting) Poincar\'e patch. This result was obtained in the approximation that $ \tau\gg \tilde \tau_0 $.
This loop contribution does not seem to simply resum to a shift in the mass and 
may be a signal of more interesting dynamics. 
In fact, Krotov and Polyakov argue for a decay of vacuum energy along the line of \cite{Polyakov:2007mm, Polyakov:2009nq} (see also\cite{Tsamis:1996qq,Tsamis:1996qm} and \cite{Mottola:1984ar, Antoniadis:2006wq, Mottola:2010gp} for similar claims). 
We however note that the result of 
\cite{Krotov:2010ma} is only valid for
$\tau \gg \tilde\tau_0$ which  
appears to imply that there is no  large logarithm in this case.
Recently, Marolf and Morrison \cite{Marolf:2011sh} have examined the question of what happens at late time and 
 they found that the full propagator approaches the Euclidean one. 
This is in agreement with the proof of IR stability on the sphere to all order by Hollands
\cite{Hollands:2010pr} and Marolf and Morrison \cite{Marolf:2010nz} (see \cite{Hollands:2011we} for a discussion of the massless case). 
It would be interesting to see if our methods could be applied to this
case to find a better result for the propagator.

Finally, we gave an explicit formula for the decay rate. It is suppressed by the Boltzman factor at large mass $\Gamma \rightarrow \lambda^2 e^{-\pi m/H}$ but it reaches a finite value of order $\Gamma \sim  \lambda^2 $ when $ m = \frac{d H}{2}$. This is in agreement with the tree level results of \cite{Bros:2006gs}.  It is tempting to interpret this decay as coming from the ``thermal" 
nature of de Sitter. In this interpretation, fields with $m> d/2 H$ have a Compton wavelength 
smaller than a Hubble patch and one can use the static coordinate to describe them. In this 
system of coordinates, the Bunch-Davies vacua looks thermal and one can naively steal energy 
from the heat bath to open new decay channels. This intuition is not borne out by 
analog computation in thermal field theory (see \cite{lebellac}). While there is in 
general an imaginary part to the self-energy at finite temperature and energy can be 
taken from heat bath, the process of a massive fields with $m\gg T$ 
 self-duplicating themselves does not occur. 

We reiterate that this imaginary part only shows up when the mass goes above the upper bound 
\be
m^2 < \frac{d^2 H^2}{4}\,.
\ee
Below this bound the self-energy is real and there is no decay rate (see for example  \cite{vanderMeulen:2007ah}).
This bound is of course very similar to the well-known Breitenlohner-Freedman \cite{Breitenlohner:1982bm, Breitenlohner:1982jf} bound in Anti-de Sitter $m^2R_{AdS}^2 < - \frac{d^2}{4}$. The de Sitter analog was proposed by Skenderis and Townsend \cite{Skenderis:2006fb} using arguments from fake supersymmetry. When one attempts to define a holographic dS/CFT correspondence (see for example \cite{Strominger:2001pn,Maldacena:2002vr}
and references therein), fields with mass above this bound are the
``harder" case to interpret. The conformal weight of these fields
living on the boundary are imaginary pointing to a non-unitary
CFT.  Here, we see that when interactions are turned on, the fields now
decay even faster at large distances. It would be interesting to work
out the implications of this decay rate for dS/CFT. 

Finally, beside these very theoretical questions, this decay rate could have practical 
implications for the phenomenology of inflationary theories. Fields with $m \sim H$ 
could be present
during inflation and they could have an effect on the spectrum of perturbations. For example, 
in the model of quasi-single field inflation \cite{Chen:2009we}, oscillations of a massive field 
lead to features in the power spectrum. In  \cite{Chen:2009we}, the authors restricted themselves
 to $m\sim H$ but still less than $\frac32 H$ and it would be interesting to see what changes 
  if the mass is above this bound. Another change to the physics may occur when 
  integrating out massive fields to get the effective action at the inflationary scale. This is usually done as in flat space, or at the very least neglecting interactions (see  
\cite{Jackson:2011qg} for a recent analysis). Integrating out massive fields with this new modified propagator may affect the quantum corrections in the effective action. We hope to return to these questions
in future work.

\hspace{1in}

{\bf Acknowledgments:}  We would like to thank Niayesh Afshordi, Emil Akhmedov, Rich Holman, Janet Hung, Sarah Shandera and Yanwen Shang for discussions.
DPJ and AR would like to thank Perimeter Institute, Waterloo for hospitality in
the initial stages of this work.  DPJ also thanks CHEP, IISc,
Bangalore for hospitality. Research at the Perimeter Institute is supported in part by the Government of Canada through NSERC and by the Province of Ontario through the Ministry of Research and Information (MRI). AR is supported in part by NSF Grant No.~PHY--0653656.

\appendix
\section{Sunset diagrams in $\lambda_4 \phi^4$}

The imaginary contribution to the self-energy is not peculiar to the cubic interaction. Here we show that for a quartic interaction of the form $\lambda_4\phi^4$, there is an imaginary component as well. (The subscript is just to differentiate the quartic coupling from the cubic coupling used in the main text). 
\begin{figure}[h]
\centering{\includegraphics[width=0.4\linewidth]{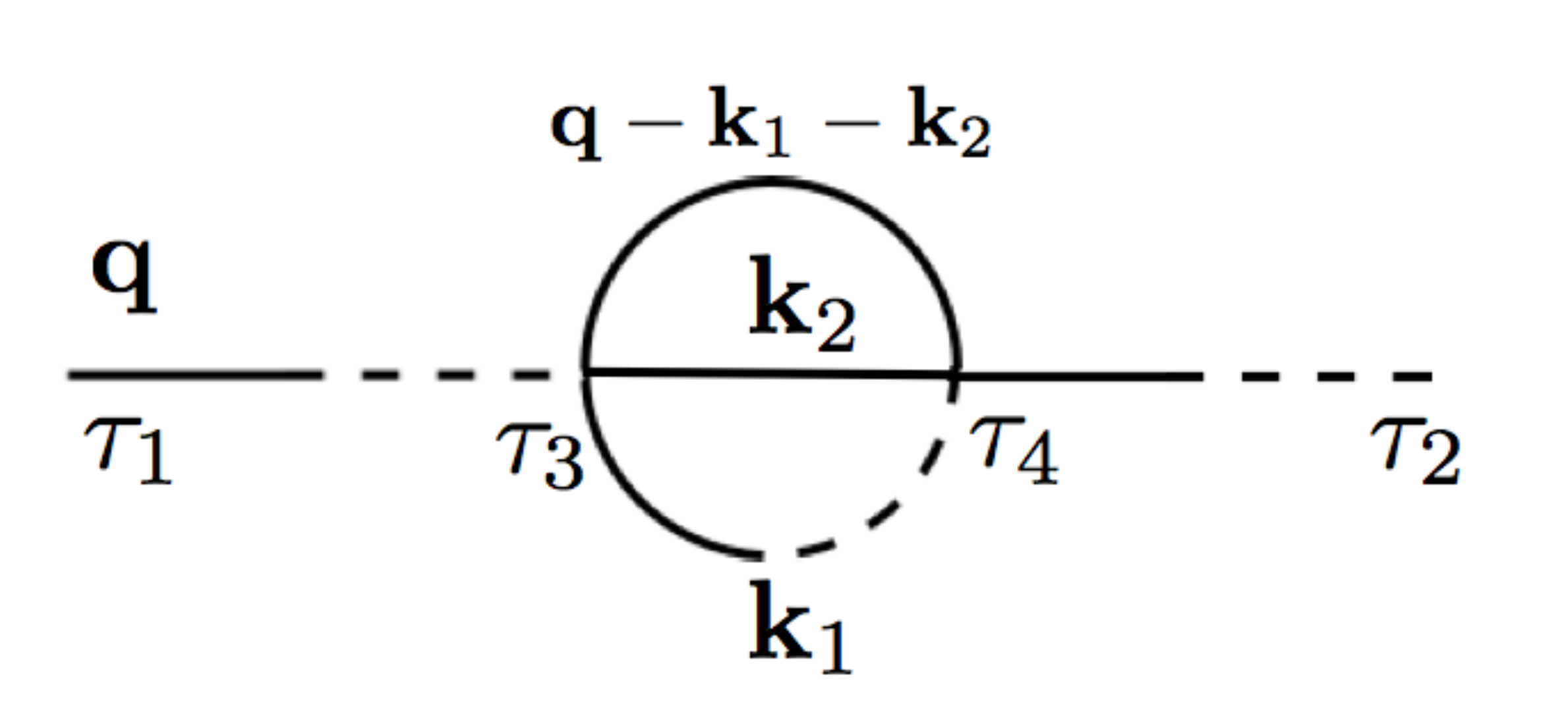}}
\caption{2-loop correction to the retarded propagator in $\lambda\phi^4$. }
\label{sunset}
\end{figure}
The relevant diagram is the one depicted in Fig.~(\ref{sunset}). This is the only sunset diagram at 2-loop for the retarded propagator. It has a very similar form to the cubic case
\bea
g^{R1} &\approx&
\lambda_4^2\int_q^{\infty} d^dk'_1d^dk'_2 \int_{\tau_2}^{\tau_1}\frac{d\tau_3}{\tau_3}
\int_{\tau_2}^{\tau_3}\frac{d\tau_4}{\tau_4}(\tau_3\tau_4)^{d}g^{R0}(q\tau_1,q\tau_3)g^{R0}(k'_1\tau_3, k_1\tau_4)\nn \\
&&\hspace{1.5in} \times f^0(k'_2\tau_3,k_2\tau_4)f^0(|\vec{k}'_1+\vec{k}'_2|\tau_3,|\vec{k}'_1+\vec{k}'_2|\tau_4)g^{R0}(q \tau_4,q\tau_2)\,,\nonumber
\eea
where we have taking $k_{1,2}\gg q$ as before. Rescaling times in terms of $k_{1,2}$ as we did in section \S~\ref{explicit} is not particularly useful here because of the norm $|\vec{k}'_1+\vec{k}'_2|$. Instead, it is easier to rescale the momenta
\be
\vec{k}_{1,2} \rightarrow \frac{\vec{k}'_{1,2}}{\sqrt{\tau_3\tau_4}} \; .
\ee
We expand the two external retarded Green's function and we keep only the interference term which contain ratios of $\tau_3/\tau_4$. As before the remaining time integral is antisymmetric under the exchange of time and so the end result is 
 \bea
g^{R1}& \sim &\lambda_4^2 \left(\frac{i}{\mu}\right) ^2 \left(
 \left(\frac{\tau_1}{\tau_2}\right)^{i\mu} -  \left(\frac{\tau_2}{\tau_1}\right)^{i\mu} \right)
 \int_{\tau_2}^{\tau_1}\frac{d\tau_3}{\tau_3}
\int_{\tau_2}^{\tau_3}\frac{d\tau_4}{\tau_4} \int_q^\infty d^d k_1d^d k_2\\
&&\hspace{1.5in} \times g^{R0}(k_1,\tau_3/\tau_4)f^0(k_2,\tau_3/\tau_4)f^0(|\vec{k}_1+\vec{k}_2|,\tau_3/\tau_4)\,,\nonumber
\eea
where we use a shortand notation $g^{R0}(k_1,\tau_3/\tau_4) = g^{R0}\left(k_1 \sqrt{\frac{\tau_3}{\tau_4}},k_1 \sqrt{\frac{\tau_4}{\tau_3}}\right)$. The integral over $k_{1,2}$ will lead to a function of $\tau_3/\tau_4$ only; call this function $\Phi$.  The integral over time is logarithmically divergent
\be 
 \int_{\tau_2}^{\tau_1}\frac{d\tau_3}{\tau_3}
\int_{\tau_2}^{\tau_3}\frac{d\tau_4}{\tau_4}  \Phi(\tau_3/\tau_4) = \int_{\tau_2}^{\tau_1} \frac{d\xi}{\xi} \int_1^{\tau_1/\tau_2} \frac{dz}{z} \Phi(z) \sim \ln \frac{\tau_1}{\tau_2}\; .
\ee
Finally the integral over momenta is imaginary because $g^R$ is imaginary while $f$ is real. In the end, the sunset diagram goes like
\be\label{phi4}
g^{R1}\sim g^{R0} \lambda_4^2 \frac{i}{\mu} I(\mu)\ln \frac{\tau_1}{\tau_2}\; ,
\ee
where $I(\mu) $ is imaginary and equal to 
\be
I(\mu) = \int_1^{\tau_1/\tau_2} \frac{dz}{z} z^{i\mu}\int_q^\infty d^d k_1d^d k_2 g^{R0}(k_1,z)f^0(k_2,z)f^0(|\vec{k}_1+\vec{k}_2|,z)\,.
\ee
The 2-loop sunset diagrams in $\lambda_4 \phi^4$ has exactly the same structure has the cubic diagram of Fig.~(\ref{loopphi3}). We can solve the Schwinger-Dyson equation consistently using exactly the same ansatz as we did and the end result is an imaginary shift to the mass. This shows that the decay rate is not particular to the cubic case with its potential unbounded from below.

\end{document}